# Thermally-robust spatiotemporal parallel reservoir computing by frequency filtering in frustrated magnets

Kaito Kobayashi*, Yukitoshi Motome

Department of Applied Physics, University of Tokyo,
Bunkyo-ku, Tokyo, 113-8656, Japan

*Corresponding author. E-mail: kaito-kobayashi92@g.ecc.u-tokyo.ac.jp

**Abstract**

Physical reservoir computing is a framework for brain-inspired information processing that utilizes nonlinear and high-dimensional dynamics in non-von-Neumann systems. In recent years, spintronic devices have been proposed for use as physical reservoirs, but their practical application remains a major challenge, mainly because thermal noise prevents them from retaining short-term memory, the essence of neuromorphic computing. Here, we propose a framework for spintronic physical reservoirs that exploits frequency domain dynamics in interacting spins. Through the effective use of frequency filters, we demonstrate, for a model of frustrated magnets, both robustness to thermal fluctuations and feasibility of frequency division multiplexing. This scheme can be coupled with parallelization in spatial domain even down to the level of a single spin, yielding a vast number of spatiotemporal computational units. Furthermore, the nonlinearity via the exchange interaction allows information processing among different frequency threads. Our findings establish a design principle for high-performance spintronic reservoirs with the potential for highly integrated devices.

**Introduction**

   Physical reservoir computing is attracting interdisciplinary attentions as one of the key technologies for realizing real-time information processing required in the coming Internet of Things society, while breaking free from energy-consuming silicon-based devices (1-3). Taking advantage of nonlinear phenomena in a physical system known as a "physical reservoir", the input data is mapped onto a high-dimensional feature space in a nonlinear manner. In addition to these two characteristics, fading memory property is considered another important feature for processing time-varying input streams (4, 5). In this framework, internal states of the reservoir are represented by a read-out function in the form of vectors, and a linear transformation of these vectors provides the final output for machine learning problems, such as classification and predictive analytics. In contrast to usual neural networks, training is limited to linear regression only at the read-out layer, which enables physical reservoirs to operate at low power, high speed, and high versatility, as demonstrated in various implementations, including photonic systems (6-14) and electrical systems (15-17). Magnetic materials are also a potential platform for physical reservoir computing, which has been proposed in various setups, such as spin torque oscillators (18-22), spin wave devices (23-25), skyrmion fabrics (25-29) and magnon scatterings (30).

   Toward device applications of spintronic reservoirs in realistic systems, two major challenges remain to be resolved. One is robustness against thermal fluctuations, which is crucial for retaining short-term memory (STM) long enough for practical use. Most spintronic reservoirs exploit nonlinear phenomena originating from magnetization dynamics, and are therefore inherently vulnerable to external noises that disturb spin precessions. The other challenge is designing devices with highly integrated computational units, the importance of which is evident from the problems facing silicon chips. The straightforward solution for the latter is the use of multiple read-outs (19, 20, 28, 31, 32), in which multiple internal state vectors are extracted from several local measurements, such as the dynamics of particular spins, and each is used independently for different computations. Another solution is parallelization by a selective read-out from superposed signals, such as wavelength-multiplexing (10, 13) and frequency-multiplexing (14) studied in photonic systems, but such schemes have been largely unexplored for spintronic reservoirs thus far.

In this article, we propose a framework of spintronic physical reservoir computing that achieves both thermal-robustness and high-integration. We demonstrate it by utilizing a simple model of frustrated magnets as a physical reservoir, with the input by AC magnetic fields and read-out by spin dynamics. We find that the memories of input information are stored in the spin dynamics at frequencies around that of the AC field and can be preserved against thermal noise by filtering out irrelevant signals at the other frequencies. We also demonstrate parallel processing on multiple inputs in both space and time, using read-outs on different spins at different frequencies. Moreover, we show that the nonlinearity arising from the exchange interaction allows calculations that capture information across various frequencies without the need for dedicated communication channels. Our results pave the way for spintronic physical reservoir computing in realistic situations.

**Results**

**Physical reservoir and input/output**

Our spintronic reservoir receives a random sequence of input binary digits $\{s_k\}$ in the form of local magnetic fields, and transforms them nonlinearly through the spin dynamics (Fig. 1a). As a prototypical model of frustrated magnets, we consider an antiferromagnetic Heisenberg model with classical spins on an $L \times L$ ($L = 128$) triangular lattice with open boundary conditions. The Hamiltonian is given by

$$\mathcal{H} = J \sum_{\langle i,j \rangle} \mathbf{S}_i \cdot \mathbf{S}_j - \sum_{j \in \Lambda} H_j^z(t) S_j^z \qquad (1)$$

where $\mathbf{S}_i = (S_i^x, S_i^y, S_i^z)$ represents the classical spin at site $i$ ($|\mathbf{S}_i| = 1$) and the sum of $\langle i,j \rangle$ is taken for nearest-neighbor sites; $H_j^z(t)$ is the time-dependent magnetic field for the input (Fig. 1b) and $\Lambda$ is a set of lattice sites used as the input terminals for the physical reservoir (Fig. 1c). Hereafter we take $J = 1$ as the energy unit. In the absence of magnetic fields, the spin configuration of this system has a 120° structure with three sublattices at low temperature (33, 34), which might be realized experimentally, for instance, in $A$CrO$_2$ ($A$ = Li, Na) (35-38), Ba$_3$CoSb$_2$O$_9$ (39, 40), and κ-(BEDT-TTF)$_2$Cu$_2$(CN)$_3$ (41, 42).

The input is a time series information of the binary bit $s_k$ with an interval of time, which we take $t_{\text{in}} = 12$, and converted into the magnetic field as

$$H_{j \in \Lambda}^z(t) = \begin{cases} H_{\text{in}} \sin(2\pi f_{\text{in}} t) & (s_k = 1) \\ -H_{\text{in}} \sin(2\pi f_{\text{in}} t) & (s_k = 0), \end{cases} \qquad (2)$$

where $H_{\text{in}}$ and $f_{\text{in}}$ is the norm and frequency of the magnetic field, respectively (Fig. 1b). We take $f_{\text{in}} = n/(2t_{\text{in}})$ with an integer $n$ so that $H_j^z(t)$ varies continuously in time. The magnetization dynamics under the input field is simulated by the stochastic Landau-Lifshitz-Gilbert (LLG) equation

$$\frac{d\mathbf{S}_i}{dt} = -\frac{1}{1+\alpha^2} \left[ \mathbf{S}_i \times \mathbf{H}_i^{\text{eff}} + \mathbf{S}_i \times \left( \mathbf{S}_i \times \mathbf{H}_i^{\text{eff}} \right) \right], \qquad (3)$$

where $\alpha$ is the Gilbert damping constant and $\mathbf{H}_i^{\text{eff}}$ is the effective magnetic field at site $i$ consisting of the input magnetic field, the exchange magnetic field, and the thermal field at temperature $T$ (43, 44) (see the Methods section). We take $\alpha = 0.1$ in the following calculations. The time profiles of the $z$ components of the read-out spins $S_j^z$ are observed every $\Delta t_{\text{in}} = t_{\text{in}}/N_t$ (Fig. 1d) where we take $N_t = 120$ (45). Then, the internal states of the reservoir, $\mathbf{X}_k$, are defined by an $(N_t + 1)$-dimensional vector including an additional component with the constant value 1 as

$$\mathbf{X}_k = \left(S_j^z(kt_{\text{in}}), S_j^z(kt_{\text{in}} + \Delta t_{\text{in}}), S_j^z(kt_{\text{in}} + 2\Delta t_{\text{in}}), \ldots, S_j^z(kt_{\text{in}} + (N_t - 1)\Delta t_{\text{in}}), 1\right). \quad (4)$$

The final output is obtained by linear transformation of the internal state vector $y_k = \mathbf{X}_k \cdot \mathbf{w}$ where $\mathbf{w}$ is a weight vector. With sufficiently large training data, the weight $\mathbf{w}$ is trained so that each $y_k$ becomes close to the desired output $\bar{y}_k$ for a given problem. The reservoir performance is evaluated by the determination coefficient $R^2$ between the sequence of outputs $\mathbf{y} = \{y_k\}$ and targets $\bar{\mathbf{y}} = \{\bar{y}_k\}$: $R^2 = \text{cov}^2(\mathbf{y}, \bar{\mathbf{y}})/[\sigma^2(\mathbf{y})\sigma^2(\bar{\mathbf{y}})]$ where cov and $\sigma^2$ denote covariance and variance, respectively. $R^2$ is close to 1 when $\mathbf{y}$ and $\bar{\mathbf{y}}$ are well matched, and approaches 0 otherwise (see the Methods section).

**Thermal robustness by frequency domain filtering**

First, we examine the STM task to evaluate how long the input information is retained in the reservoir. The target output for this task is $\bar{y}_k = s_{k-d}$ where $d$ is the delay step after the input. Here, we apply the input magnetic field at frequency $f_{\text{in}} = 8/t_{\text{in}}$ to one specific terminal spin at site $j$, namely, $\Lambda = \{j\}$, and use the spin dynamics at the same site, $S_j^z(t)$, as a read-out to construct the internal state vectors. We place $j$ around the center of the system to avoid uninteresting behavior near the edges.

Figure 2a shows temperature dependence of the reservoir performance for the STM task. At absolute zero temperature, the memories are recovered with high accuracy of $R^2 > 0.95$ up to delay $d = 7$, and decay gradually for larger $d$, exhibiting the fading memory property required for a physical reservoir. In contrast, at finite temperature, the reservoir rapidly loses the STM older than $d = 2$ even at extremely low temperature $T = 10^{-4}$, and furthermore, remaining memories of $d = 0$ and 1 are also lost as the temperature increases. This is an indication that the STM stored in the spin dynamics is extremely fragile against thermal noise, as commonly seen in various spintronic physical reservoirs (18-28).

To clarify how the thermal noise disturbs the read-out signals, we analyze the Fourier spectrum of the dynamics of $S_j^z(t)$. Figure 2d displays the spectra at three different temperatures. At $T = 0$, the spectrum shows strong peaks at around the frequency of the input magnetic field, $\pm f_{\text{in}}$, as the terminal spin linearly follows the time-dependent magnetic field via the Zeeman coupling. The wavy structure repeating every $1/(2t_{\text{in}})$ also appears, which originates from the random switching of the input magnetic field at every $t_{\text{in}}$ according to the input bit. Here, the finite peak width at $\pm f_{\text{in}}$ reflects the probabilistic nature of the bit sequence $\{s_k\}$ in the input magnetic field, otherwise there would be only a delta-functional peak at exactly $\pm f_{\text{in}}$. In contrast, at finite temperature, thermal noise disturbs the spin dynamics, obscuring these characteristic peaks. At $T = 0.001$, the wavy structure disappears due to the thermal noise, but the peaks at around $\pm f_{\text{in}}$ are still present. The latter peaks are also mostly drowned out at a higher temperature, $T = 0.1$.

From this observation, we introduce a frequency filter in order to mitigate the disturbances caused by thermal agitations. Specifically, we only retain signals within the frequency windows of $f_{\text{in}} \leq |f| < f_{\text{in}} + 1/(2t_{\text{in}})$, and use the frequency-filtered spin dynamics $S_{j,\text{filtered}}^z(t)$ for computation instead of the whole spin dynamics, $S_j^z(t)$; see Fig. 2c and the Methods section. Figure 2b shows the reservoir performance for the STM task with the frequency filter. In clear contrast to the case without frequency filtering in Fig. 2a, the STM persists up to $d = 5$ with almost no information loss even at finite temperature, and the older memories are lost gradually. This indicates, surprisingly, that the input information can be recovered only by the signals within the narrow frequency bandwidth of $1/(2t_{\text{in}})$ including $f_{\text{in}}$, and that the memories within this range are resistant to thermal fluctuations.

Figure 2e shows the distributions of the STM in frequency domain. We introduce frequency filters that pass signals within the frequency window of $(m-1)/(2t_{\text{in}}) \leq |f| < m/(2t_{\text{in}})$, and evaluate the performance for the STM task while changing a positive integer $m$. At $T = 0$, the memories are encoded within a broad frequency range, but $R^2$ gradually diminishes as the frequency window deviates from $f_{\text{in}}$. Since the information embedding process from the input magnetic field is mainly governed by the linear Zeeman interaction, the input information is prominently restored in the spin dynamics at around $f_{\text{in}}$. At $T = 0.001$, the system performance at around $f_{\text{in}}$ remains largely unchanged from that at $T = 0$, although it deteriorates significantly at other frequencies. A comparison with Fig. 2d demonstrates that the input information is retained only at frequencies where the spin dynamics is preserved against thermal noise. As the signals at the other frequencies contribute as noises, the reservoir without the frequency filter is vulnerable to thermal agitations. At $T = 0.1$, the reservoir is almost non-functional even at around $f_{\text{in}}$ because of the increased noise level as shown in Fig. 2b. In our configuration, we employ a single spin as the input and read-out, however, we verify that the utilization of multiple spins or the entire system does not alter the fundamental behavior. We note that increasing the amplitude of the input magnetic field broadens the temperature range within which the system can operate effectively (see Supplementary Note 1 and Supplementary Fig. 1).

The influence of the input information in frequency domain is closely tied to the typical time scale of the system. In our input protocol, the input magnetic field is decomposed into the product of an AC magnetic field at frequency $f_{\text{in}}$ and a pulse square field that randomly changes its sign every $t_{\text{in}}$ depending on the supplied bit. This switching duration is the only relevant time scale for the randomness of the input, and thus the information is stored in units of $1/t_{\text{in}}$ in frequency domain. In other words, the memories retained at around $f$, $f \pm n/t_{\text{in}}$ and $-f + n/t_{\text{in}}$ with a positive interger $n$ are almost identical (see Supplementary

Note 2 and Supplementary Fig. 2). This observation suggests that filters of $(m-1)/(2t_{\text{in}}) \leq |f| < m/(2t_{\text{in}})$ are the narrowest frequency bandwidth that enables maximum recovery of STM without overlapping information.

**Parallel reservoir computing in frequency and spatial domains**

The above concept of frequency filtering can be extended to implement frequency division multiplexing, which allows for the simultaneous transmission of multiple signals encoded at different frequencies. As an illustration, we prepare three sequences of random input bits $\{s_k^1\}, \{s_k^2\}, \{s_k^3\}$ and examine the STM task of each input bit: the target output is $\bar{y}_k^n = s_{k-d}^n$ with delay step $d$ and $n = 1, 2, 3$. We choose different frequencies $f_{\text{in}}^1 = 8/t_{\text{in}}$, $f_{\text{in}}^2 = 45/(2t_{\text{in}})$, $f_{\text{in}}^3 = 36/t_{\text{in}}$ for each binary sequence, and transform the bits $s_k^n$ to the magnetic field at the assigned frequency in the same manner as Eq. 2: superposition of these three magnetic fields is given to one input terminal spin at site $j$ (Fig. 3a). Herein the input bits $s_k^n$ are nonlinearly transformed through the dynamics of the terminal spin $S_j^z(t)$ and the internal state vectors $X_k$ are then extracted for computations. Figure 3b shows the Fourier spectrum of $S_j^z(t)$ at zero temperature, where three peaks corresponding to $\pm f_{\text{in}}^1, \pm f_{\text{in}}^2, \pm f_{\text{in}}^3$ appear in addition to the wavy structure present over a wide frequency range.

Figure 3c represents the reservoir performance for the STM tasks of three input bit sequences in frequency domain. Each information is embedded into the spin dynamics at around its input frequency and not affected by the other input bits. Consequently, the memories can be selectively recovered using a frequency filter that passes the input frequency where the desired bits are provided. Indeed, for all $n$, $R^2$ at around each $f_{\text{in}}^n$ is greater than 0.9 up to delay $d = 5$ and gradually decreases for larger $d$ due to the fading memory property, which indicates that the performance is almost the same for different parallel threads. This clearly demonstrates that the spin dynamics at different frequencies can be regarded as mutually independent parallel threads, allowing for frequency division multiplexed computational units. This parallelization scheme works at finite temperature in the same manner as Fig. 2e by filtering out irrelevant signals. Worth noting that the degree of parallelization can be further increased by allocating more frequencies for computation; we confirm that a spacing of $1/t_{\text{in}}$ or more is sufficient for such parallelization.

Our reservoir can also be parallelized in spatial domain with multiple inputs and read-outs. To demonstrate this, we perform the STM task by selecting two input terminal spins at sites $j_1$ and $j_2$ separated by 17 spins, namely $\Lambda = \{j_1, j_2\}$ in Eq. 1, denoted as terminal 1 and terminal 2, respectively (Fig. 4a). A bit sequence of $\{s_k^{j_1,1}\}$ ($\{s_k^{j_1,2}\}$) is given to terminal 1 and $\{s_k^{j_2,1}\}$ ($\{s_k^{j_2,2}\}$) is given to terminal 2 at frequency $f_{\text{in}}^1$ ($f_{\text{in}}^2$); in total, four input bit sequences are simultaneously supplied to the system. Figure 4b shows the performance for the STM tasks of different input sequences at zero temperature. For computation, the signals received from terminal 1 and terminal 2 are independently utilized with frequency filters. When the spin dynamics at terminal 1 is used as the read-out, the memories of $\{s_k^{j_1,1}\}$ and $\{s_k^{j_1,2}\}$ can be reproduced at around each input frequency with $R^2 > 0.9$ up to delay $d = 5$, whereas the information of $\{s_k^{j_2,1}\}$ and $\{s_k^{j_2,2}\}$ cannot be recovered at all. The opposite is true when utilizing signals from terminal 2. This illustrates that the spin dynamics at different sites can be treated as independent parallel computational threads, which realizes spatial multiplexing of computational units while preserving the frequency multiplexing. Due to the availability of single spins, the minimal components of magnetic materials, as read-outs, our proposed scheme offers the potential for high-density integration at the nanoscale on one magnetic chip; we confirm that a lattice spacing more than 10 spins is sufficient for such parallel computation in our model. Moreover, simultaneous parallelization in spatial and frequency domains would allow for extreme integration of computational units in spatiotemporal domain by further increasing the number of frequency bands and I/O terminal spins.

**Logic gate operation of bits in different threads**

Applying the characteristics of frequency division multiplexing, different information retained in parallel at different frequencies can be accessed simultaneously by selecting signals at appropriate frequencies. To demonstrate this, we examine logic gate tasks between input bits in different parallel threads (9, 29). Two binary sequences $\{s_k^1\}$ and $\{s_k^2\}$ are simultaneously given to one input terminal spin at site $j$ encoded by the frequency $f_{\text{in}}^1 = 8/t_{\text{in}}$ and $f_{\text{in}}^2 = 45/(2t_{\text{in}})$, respectively. Here, we observe the spin dynamics on the nearest neighbor site from the input terminal site, denoted as $j^{\text{NN}}$, for the read-out from the reservoir; the spin dynamics at the terminal site $j$ is also traced for a comparison (Fig. 5a). The target output $\bar{y}_k$ is the bit where two bits $s_k^1$ and $s_k^2$ are processed by a logic gate, the truth table of which is summarized in Fig. 5b. Figure 5c represents the Fourier spectra of the spin dynamics on the terminal site $j$ and the neighboring site $j^{\text{NN}}$ at zero temperature. While the overall behaviors, especially strong peaks at $\pm f_{\text{in}}^1$ and $\pm f_{\text{in}}^2$, are common between the two spectra, a significant difference is found in additional peaks in the spectrum from the site $j^{\text{NN}}$ at around the second harmonics frequencies: $\pm 2f_{\text{in}}^1$, $\pm 2f_{\text{in}}^2$, $\pm(f_{\text{in}}^2 - f_{\text{in}}^1)$ and $\pm(f_{\text{in}}^1 + f_{\text{in}}^2)$. The signals at $\pm 2f_{\text{in}}^1$ and $\pm 2f_{\text{in}}^2$, which originate from the square of magnetic fields at the same frequency, are delta-functional since the sign information of binary bits is lost. In contrast, the signals at $\pm(f_{\text{in}}^2 - f_{\text{in}}^1)$ and $\pm(f_{\text{in}}^1 + f_{\text{in}}^2)$ resulting from nonlinear process at different frequencies have finite widths because the randomness of input bits is preserved.

To begin with, the logic gate tasks — AND, OR, XOR, NAND, NOR and XNOR — are demonstrated using signals in the frequency windows of $f_{\text{in}}^1 - 1/(2t_{\text{in}}) \leq |f| < f_{\text{in}}^1 +$ $1/(2t_{in})$ and $f_{in}^2 - 1/(2t_{in}) \leq |f| < f_{in}^2 + 1/(2t_{in})$. Figure 5d plots the reservoir performance for each task, where the blue and red bars exhibit $R^2$ with read-outs on the input site $j$ and the nearest neighbor site $j^{NN}$. By utilizing signals from the input site, the reservoir shows moderate performance with $R^2 \sim 0.6$ for linearly separable tasks, namely AND, OR, NAND, and NOR; conversely, the performance for linearly inseparable tasks, XOR and XNOR, is extremely poor with $R^2 \sim 0$. When utilizing the nearest neighbor spin for the read-out, the performance becomes slightly better, but $R^2$ for the inseparable tasks are still low. Thus, although the signals containing input frequencies of $\pm f_{in}^1$ and $\pm f_{in}^2$ should possess information of both $s_k^1$ and $s_k^2$ as discussed above, the memories are concentrated around their assigned frequencies and have little effects on one another, making it difficult to resolve tasks that are heavily reliant on information in both frequencies.

In contrast, the spin dynamics at around the frequencies of $\pm(f_{in}^2 - f_{in}^1)$ and $\pm(f_{in}^1 + f_{in}^2)$ are equally influenced by $\pm f_{in}^1$ and $\pm f_{in}^2$ via the nonlinear process, retaining the memories of both $s_k^1$ and $s_k^2$. This is demonstrated by utilizing the spin dynamics at the neighboring site $j^{NN}$ as a read-out, where information is nonlinearly transferred from the input site $j$ via the exchange interaction. Figure 5e exhibits the reservoir performance for logic gate tasks using signals in the frequency windows of $f_{in}^2 - f_{in}^1 - 1/(2t_{in}) \leq |f| < f_{in}^2 - f_{in}^1 + 1/(2t_{in})$ and $f_{in}^1 + f_{in}^2 - 1/(2t_{in}) \leq |f| < f_{in}^1 + f_{in}^2 + 1/(2t_{in})$. With read-out on the nearest neighbor site, the reservoir operates very well with $R^2 > 0.9$ for both linearly separable and inseparable tasks. These findings clearly indicate that the nonlinear processes via the exchange interaction enable two bits in different threads to exert an influence on one another. The importance of this coupling is evident from the very low $R^2$ when utilizing the input spin as the read-out in Fig. 5e, where the dynamics is dominated by the linear Zeeman coupling. Multi-bit tasks such as gating require mixing different information within the same reservoir dynamics, otherwise different bits cannot establish a nonlinear relationship. This is thus an operation that takes full advantage of the principle of superposition in the frequency domain, and is almost impossible with other parallelization schemes such as spatial domain multiplexing. We note, however, that the second harmonics in the dynamics of the nearest neighbor spin are vulnerable to thermal noise, making linearly inseparable gating almost infeasible at finite temperature. This issue would be resolved by enhancing nonlinearity in models with different interactions, as we shall discuss later.

**Discussion**

In this paper, we have proposed a framework for a physical reservoir computing in magnetic materials utilizing frequency domain dynamics. The input through an AC magnetic field and the selective read-out with frequency filters enable both thermal noise reduction and frequency division multiplexing, which are of great practical significance in the design of spintronic physical reservoir computing devices. In addition, by utilizing multiple read-outs, a substantial number of computational units can be integrated through parallelization in spatiotemporal domain. The inherent nonlinearity in the interacting spin systems allows mixing of information stored in the spin dynamics at different frequencies.

In magnetic materials, the linearly introduced information via the Zeeman coupling is nonlinearly processed by the exchange interaction. This is where the differences in materials are most pronounced. For instance, an increase in connectivity and complexity of interactions would enhance the nonlinearity of the spin dynamics, and longer-ranged interactions would propagate the input information over a wider region. When electrons in the system exhibit itinerant nature rather than localized one, the synergy between charge and spin degrees of freedom might alter the information transfer and processing. Moreover, in quantum spin systems, the reservoir takes advantage of the large dimensionality of Hilbert space, where quantum correlations and quantum entanglement would strongly influence the performance of the reservoir (31, 32, 46, 47). The impact of specific interaction on information processing will be a subject of future research.

The complexity of the magnetic structure contributes to the high dimensionality of the feature space where information is projected by a physical reservoir, providing ample resources for machine learning tasks. Here, the dimension corresponds to the number of spins whose dynamics is linearly independent, which determines the number of parameters in the feature space. In general, the spin dynamics in a uniform environment, such as a ferromagnetic state, can be described with a few parameters, while those in a complex environment require many parameters. Our reservoir with a frustrated magnet, consisting of only three sublattices, yields low dimensional internal states, yet proves sufficient for the tasks demonstrated; we confirm magnets with solely nearest-neighbor ferromagnetic interactions are too simple to be employed for computation. Physical reservoirs with more complex magnetic textures would project information onto higher dimensional feature space, which would enhance computational performance for complicated tasks, as presented in skyrmion systems (27, 28).

In addition to its thermal robustness and parallel processing capabilities, our scheme exhibits versatility in its broad applicability to a variety of spintronic physical reservoirs with input by AC signals and read-out by time-varying signals. In magnetoelectric systems, for instance, the information might be embedded and extracted via either magnetic or electrical AC signals, while taking advantage of the nonlinearity of the spin-charge coupled dynamics for information processing (48, 49). The input information should be stored within the spin and charge dynamics at the characteristic frequency bands, where frequency filtering would play an important role. Consequently, our implementation-independent protocol would be beneficial for realizing thermal robustness and frequency division multiplexing in various types of spintronic physical reservoirs. Our discovery opens up the way for hardware implementation of large-scale intelligent systems that exploits magnetization dynamics for real-time computation.

## Methods

### LLG calculation

To simulate the magnetization dynamics, we employ the stochastic LLG equation summarized in Eq. 3. The effective magnetic field includes the exchange field $\mathbf{H}_i^{\text{ex}}$ and the thermal field $\mathbf{H}_i^{\text{th}}$. The former is given by the $\mathbf{S}_i$ derivative of the Hamiltonian (Eq. 1) as $\mathbf{H}_i^{\text{ex}} = -\partial\mathcal{H}/\partial\mathbf{S}_i$, and the latter is introduced as a white noise with the following properties:

$$\langle\mathbf{H}_i^{\text{th}}(t)\rangle = 0, \qquad \langle\mathbf{H}_i^{\text{th}}(t)\mathbf{H}_j^{\text{th}}(t')\rangle = \frac{2\alpha T}{\Delta t}\delta(t-t')\delta_{ij} \qquad (5)$$

where $\Delta t$ denotes the time step (we take $\Delta t$=0.005), $\delta(t)$ the Dirac delta function, and $\delta_{ij}$ the Kronecker delta function. The time evolution is traced by the fourth-order Runge-Kutta method. At each temperature, equilibrium spin configurations are obtained by relaxation for a sufficient time starting from a complete in-plane 120° structure, and they are used as an initial state of the reservoir. The temperature dependences of the energy and specific heat are in agreement with the previous results using Monte Carlo calculations (33, 50).

### Scheme for training and testing

We take sequences of binary digits $\{s_k\}$, and each data $s_k$ is fed to the system by the out-of-plane input magnetic field to the input terminal spin. The internal state is characterized by the induced out-of-plane dynamics at the read-out spin as the $1 \times (N_t + 1)$-dimensional vector $\mathbf{X}_k$ (Eq. 4) that is then linearly transformed to the output $y_k$ by the $(N_t + 1) \times 1$-dimensional weight $\mathbf{w}$. In total $l^{\text{tot}} = 6000$ inputs are used: $l^{\text{w}} = 2000$ to warm up the system, $l^{tr} = 2000$ for training of **w**, and $l^{ts} = 2000$ for testing the reservoir performance. Since they are sufficiently large numbers compared to the number of tuning parameters $N_t + 1$ ($N_t = 120$), the generalization performance is preserved without overfitting. The vectors $\{\mathbf{X}_k\}$ at training phase and test phase are summarized to an $l^{tr} \times (N_t + 1)$ matrix $\mathbf{X}^{tr} = \{\mathbf{X}_k\}_{k=l^w}^{l^w+l^{tr}}$ and an $l^{ts} \times (N_t + 1)$ matrix $\mathbf{X}^{tr} = \{\mathbf{X}_k\}_{k=l^w+l^{tr}}^{l^w+l^{tr}+l^{ts}}$, and correspondingly the target output vectors are given by an $l^{tr} \times 1$ vector $\bar{\mathbf{y}}^{tr} = \{\bar{y}_k\}_{k=l^w}^{l^w+l^{tr}}$ and an $l^{ts} \times 1$ vector $\bar{\mathbf{y}}^{ts} = \{\bar{y}_k\}_{k=l^w+l^{tr}}^{l^w+l^{tr}+l^{ts}}$. The weight should be trained to satisfy $\bar{\mathbf{y}}^{tr} = \mathbf{X}^{tr} \cdot \mathbf{w}$, which is impossible for overdetermined systems with $l^{tr} \gg N_t + 1$. Alternatively, we optimize **w** to minimize the least squares error between vectors on both sides, which is analytically given by $\mathbf{w} = \mathbf{X}^{tr^+} \cdot \bar{\mathbf{y}}^{tr}$ where $\mathbf{X}^+$ is the Moore-Penrose pseudoinverse-matrix. The output in the testing phase is calculated by $\mathbf{y} = \mathbf{X}^{ts} \cdot \mathbf{w}$ and the similarity to the desired output $\bar{\mathbf{y}}^{ts}$ is evaluated using the determination coefficient $R^2 = \text{cov}^2(\mathbf{y}^{ts}, \bar{\mathbf{y}}^{ts})/[\sigma^2(\mathbf{y}^{ts})\sigma^2(\bar{\mathbf{y}}^{ts})]$.

**Frequency filtering**

The spin dynamics in the training phase $S^{tr}(t)$ and in the testing phase $S^{ts}(t)$ are calculated every time step $\Delta t=0.005$ by the stochastic LLG equation, with the observation time being $l^{tr} \times t_{in} = 24000$ for training and $l^{ts} \times t_{in} = 24000$ for testing. The dynamics in time domain $S^{tr}(t)$ and $S^{ts}(t)$ are independently Fourier transformed to $S^{tr}(f)$ and $S^{ts}(f)$ with a sampling frequency of $1/\Delta t = 200$ and a frequency resolution of $1/24000$. Of these, only the desired frequency components are inverse Fourier transformed to real-time dynamics $S^{tr}_{filtered}(t)$ and $S^{ts}_{filtered}(t)$, which are then used to the construct internal state vectors $\mathbf{X}^{tr}$ and $\mathbf{X}^{ts}$, respectively.

**Data availability**

The data that support the findings of this study are available upon reasonable request.

**Code availability**

The simulation codes used in this study are available upon reasonable request.

**References**

1. Roy, K., Jaiswal, A. & Panda, P. Towards spike-based machine intelligence with neuromorphic computing. *Nature* **575**, 607-617 (2019).
2. Tanaka, G. et al. Recent advances in physical reservoir computing: a review. *Neural Networks* **115**, 100-123 (2019).
3. Maass, W., Natschläger, T. & Markram, H. Real-time computing without stable states: a new framework for neural computation based on perturbations. *Neural Comput.* **14**, 2531-2560 (2002).
4. Jaeger, H. The "echo state" approach to analysing and training recurrent neural networks with an erratum note. GMD Report 148 http://www.faculty.jacobs-university.de/hjaeger/pubs/EchoStatesTechRep.pdf (German National Research Institute for Computer Science, 2001).
5. Jaeger, H. & Haas, H. Harnessing nonlinearity: predicting chaotic systems and saving energy in wireless communication. *Science* **304**, 78-80 (2004).
6. Duport, F., Schneider, B., Smerieri, A., Haelterman, M. & Massar, S. All-optical reservoir computing. *Opt. Express* **20**, 22783-22795 (2012).
7. Paquot, Y. et al. Optoelectronic reservoir computing. *Sci. Rep.* **2**, 287 (2012).

8. Brunner, D., Soriano, M. C., Mirasso, C. R., & Fischer, I. Parallel photonic information processing at gigabyte per second data rates using transient states. *Nat. Commun.* **4**, 1364 (2013).

9. Vandoorne, K. et al. Experimental demonstration of reservoir computing on a silicon photonics chip. *Nat. Commun.* **5**, 3541 (2014).

10. Tait, A. N., Nahmias, M. A., Shastri, B. J., & Prucnal, P. R. Broadcast and weight: an integrated network for scalable photonic spike processing. *J. Lightwave Technol.* **32**, 3427-3439 (2014).

11. Van der Sande, G., Brunner, D., & Soriano, M. C. Advances in photonic reservoir computing. *Nanophotonics* **6**, 561-576 (2017).

12. Shastri, B. J. et al. Photonics for artificial intelligence and neuromorphic computing. *Nat. Photonics* **15**, 102–114 (2021).

13. Zipp, L. J. & Stoker, D. S. Dual time- and wavelength-multiplexed photonic reservoir computing. in *AI and Optical Data Sciences II*, Proceedings of SPIE Vol. 11703 (SPIE, Bellingham, WA, 2021) 1170305.

14. Butschek, L. et al. Photonic reservoir computer based on frequency multiplexing. *Opt. Lett.* **47**, 782-785 (2022).

15. Du, C. et al. Reservoir computing using dynamic memristors for temporal information processing. *Nat. Commun.* **8**, 2204 (2017).

16. Zhong, Y. et al. Dynamic memristor-based reservoir computing for high-efficiency temporal signal processing. *Nat. Commun.* **12**, 408 (2021).

17. Milano, G. et al. In materia reservoir computing with a fully memristive architecture based on self-organizing nanowire networks. *Nat. Mater.* **21**, 195-202 (2022).

18. Torrejon, J. et al. Neuromorphic computing with nanoscale spintronic oscillators. *Nature* **547**, 428-431 (2017).

19. Furuta, T. et al. Macromagnetic simulation for reservoir computing utilizing spin dynamics in magnetic tunnel junctions. *Phys. Rev. Applied* **10**, 034063 (2018).

20. Kanao, T. et al. Reservoir computing on spin-torque oscillator array. *Phys. Rev. Applied* **12**, 024052 (2019).

21. Tsunegi, S. et al. Physical reservoir computing based on spin torque oscillator with forced synchronization. *Appl. Phys. Lett.* **114**, 164101 (2019).

22. Marković, D. et al. Reservoir computing with the frequency, phase, and amplitude of spin-torque nano-oscillators. *Appl. Phys. Lett.* **114**, 012409 (2019).

23. Nakane, R., Tanaka, G. & Hirose, A. Reservoir computing with spin waves excited in a garnet film. *IEEE Access* **6**, 4462-4469 (2018).

24. Nakane, R., Hirose, A. & Tanaka, G. Spin waves propagating through a stripe magnetic domain structure and their applications to reservoir computing. *Phys. Rev. Research* **3**, 033243 (2021).

25. Lee, M.-K. & Mochizuki, M. Reservoir computing with spin waves in a skyrmion crystal. *Phys. Rev. Applied* **18**, 014074 (2022).

26. Prychynenko, D. et al. Magnetic skyrmion as a nonlinear resistive element: a potential building block for reservoir computing. *Phys. Rev. Applied* **9**, 014034 (2018).

27. Pinna, D., Bourianoff, G., & Everschor-Sitte, K. Reservoir computing with random skyrmion textures. *Phys. Rev. Applied* **14**, 054020 (2020).

28. Yokouchi, T. et al. Pattern recognition with neuromorphic computing using magnetic field-induced dynamics of skyrmions. *Sci. Adv.* **8**, eabq5652 (2022).

29. Raab, K. et al. Brownian reservoir computing realized using geometrically confined skyrmion dynamics. *Nat. Commun.* **13**, 6982 (2022).

30. Körber, L. et al. Pattern recognition with a magnon-scattering reservoir. Preprint at https://arxiv.org/abs/2211.02328 (2022).

31. Nakajima, K., Fujii, K., Negoro, M., Mitarai, K. & Kitagawa, M. Boosting computational power through spatial multiplexing in quantum reservoir computing. *Phys. Rev. Applied* **11**, 034021 (2019).

32. Bravo, R. A., Najafi, K., Gao, X. & Yelin, S. F. Quantum reservoir computing using arrays of Rydberg atoms. *PRX Quantum* **3**, 030325 (2022).

33. Kawamura, H. & Miyashita, S. Phase transition of the two-dimensional Heisenberg antiferromagnet on the triangular lattice. *J. Phys. Soc. Jpn.* **53**, 4138 (1984).

34. Kawamura, H., Yamamoto, A. & Okubo, T. $Z_2$-vortex ordering of the triangular-lattice Heisenberg antiferromagnet. *J. Phys. Soc. Jpn.* **79**, 023701 (2010).

35. Olariu, A. et al. Unconventional dynamics in triangular Heisenberg antiferromagnet $NaCrO_2$. *Phys. Rev. Lett.* **97**, 167203 (2006).

36. Alexander, L. K., Büttgen, N., Nath, R., Mahajan, A. V. & Loidl, A. $^7$Li NMR studies on the triangular lattice system $LiCrO_2$. *Phys. Rev. B* **76**, 064429 (2007).

37. Hsieh, D. et al. Unconventional spin order in the triangular lattice system $NaCrO_2$: A neutron scattering study. *Physica B* **403**, 1341-1343 (2008).

38. Sugiyama, J. et al. $\mu^+$SR investigation of local magnetic order in $LiCrO_2$. *Phys. Rev. B* **79**, 184411 (2009).

39. Shirata, Y., Tanaka, H., Matsuo, A. & Kindo, K. Experimental realization of a spin-1/2 triangular-lattice Heisenberg antiferromagnet. *Phys. Rev. Lett.* **108**, 057205 (2012).

40. Kamiya, Y. et al. The nature of spin excitations in the one-third magnetization plateau phase of $Ba_3CoSb_2O_9$. *Nat. Commun.* **9**, 2666 (2022).

41. Yamashita, S. et al. Thermodynamic properties of a spin-1/2 spin-liquid state in a κ-type organic salt. *Nature Phys.* **4**, 459-462 (2008).

42. Yamashita, M. et al. Thermal-transport measurements in a quantum spin-liquid state of the frustrated triangular magnet κ-(BEDT-TTF)$_2$Cu$_2$(CN)$_3$. *Nature Phys.* **5**, 44-47 (2009).

43. Brown, W. F. Thermal fluctuations of a single-domain particle. *Phys. Rev.* **130**, 1677 (1963).

44. Leliaert, J. et al. Adaptively time stepping the stochastic Landau-Lifshitz-Gilbert equation at nonzero temperature: Implementation and validation in MuMax$^3$. *AIP Advances* **7**, 125010 (2017).

45. Appeltant, L. et al. Information processing using a single dynamical node as complex system. *Nat. Commun.* **2**, 468 (2011).

46. Fujii, K. & Nakajima, K. Harnessing disordered-ensemble quantum dynamics for machine learning. *Phys. Rev. Applied* **8**, 024030 (2017).

47. Martínez-Peña, R., Giorgi, G. L., Nokkala, J., Soriano, M. C. & Zambrini R. Dynamical phase transitions in quantum reservoir computing. *Phys. Rev. Lett.* **127**, 100502 (2021).

48. Scott, J. Multiferroic memories. *Nature Mater.* **6**, 256-257 (2007).

49. Tokura, Y., Seki, S. & Nagaosa, N. Multiferroics of spin origin. *Rep. Prog. Phys.* **77**, 076501 (2014).

50. Capriotti, L., Vaia, R., Cuccoli, A. & Tognetti, V. Phase transitions induced by easy-plane anisotropy in the classical Heisenberg antiferromagnet on a triangular lattice: A Monte Carlo simulation. *Phys. Rev. B* **58**, 273 (1998).

**Acknowledgements**

We thank K. Shimizu for fruitful discussions. This work was supported by a Grant-in-Aid for Scientific Research on Innovative Areas "Quantum Liquid Crystals" (KAKENHI Grant No. JP19H05825) from JSPS of Japan. K.K. was supported by the

**Author contributions**

**Competing interests**

# Figures

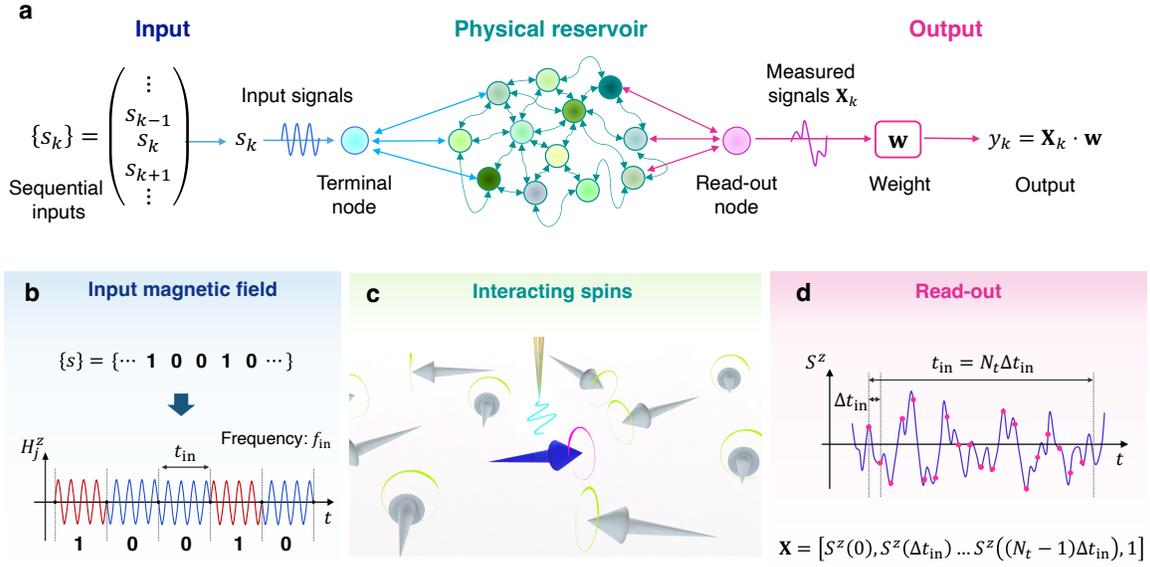

**Fig. 1: Concept of physical reservoir computing by utilizing a frustrated magnet. a** Architecture of physical reservoir computing, which consists of the input, reservoir, and output. Sequential information is converted to physical signals in the input part and nonlinearly transformed in the reservoir part via its dynamics. The output part linearly transforms measured signals from the physical reservoir to final outputs, whose coefficients are trained so that the final signals give desired answers for a specific problem. **b** Schematic diagram of the input magnetic field. The $k$-th input bit $s_k$ is converted to an AC magnetic field $H_j^z(t)$ with the duration of $t_{\text{in}}$; see Eq. 2. **c** Schematic of an antiferromagnetic Heisenberg model on a triangular lattice used as the physical reservoir. The blue arrow represents the terminal spin where the input magnetic fields are applied using a probe indicated by the yellow needle, and the gray arrows represent the other spins. The $z$ components of the spin dynamics on the terminal sites are observed as the read-out. **d** Schematic diagram of read-out from the physical reservoir. The spin component is observed

every $\Delta t_{\text{in}} = t_{\text{in}}/N_t$ and transformed to an $(N_t + 1)$-dimensional internal state vector $\mathbf{X}_k$; see Eq. 4. The final output $y_k$ is calculated as $y_k = \mathbf{X}_k \cdot \mathbf{w}$ with the weight vector $\mathbf{w}$.

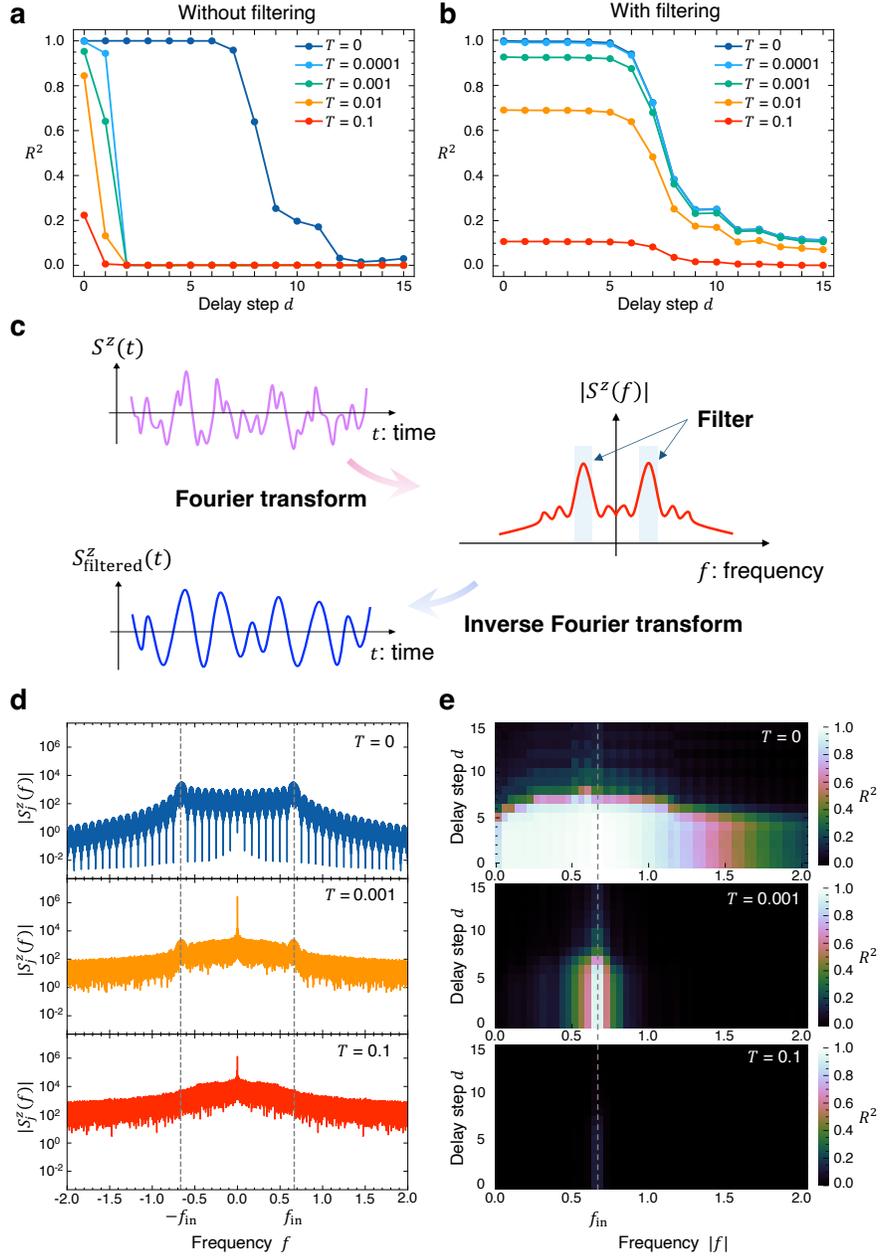

**Fig. 2: Reservoir computing in the Heisenberg antiferromagnet at finite temperature.**

**a** Reservoir performance for the STM task at several temperatures $T$ when using the whole spin dynamics (without frequency filtering). Delay step $d$ denotes discretized time length after input and $R^2$ is the determination coefficient. The STM is extremely fragile against thermal noise, especially for $d \geq 2$. **b** The same plot as **a** with the frequency filter of $f_{in} \leq$ $|f| < f_{in} + 1/(2t_{in})$. The memories are retained even at finite temperatures, proving the efficiency of frequency filtering for thermal noise. **c** Schematic picture of a frequency filter. The signals within a specific frequency band are extracted by means of the (inverse) Fourier transformation. **d** Fourier spectra of the spin dynamics on the terminal site $S_j^z(t)$ at three different temperatures. The dashed lines display the input frequency $\pm f_{in}$. **e** Distributions of the STM in frequency domain. The color represents the value of $R^2$ and the dashed line shows $f_{in}$. The memories are stored at around $f_{in}$ where the spin dynamics is preserved against thermal fluctuations, allowing for an efficient reservoir computing even at finite temperature.

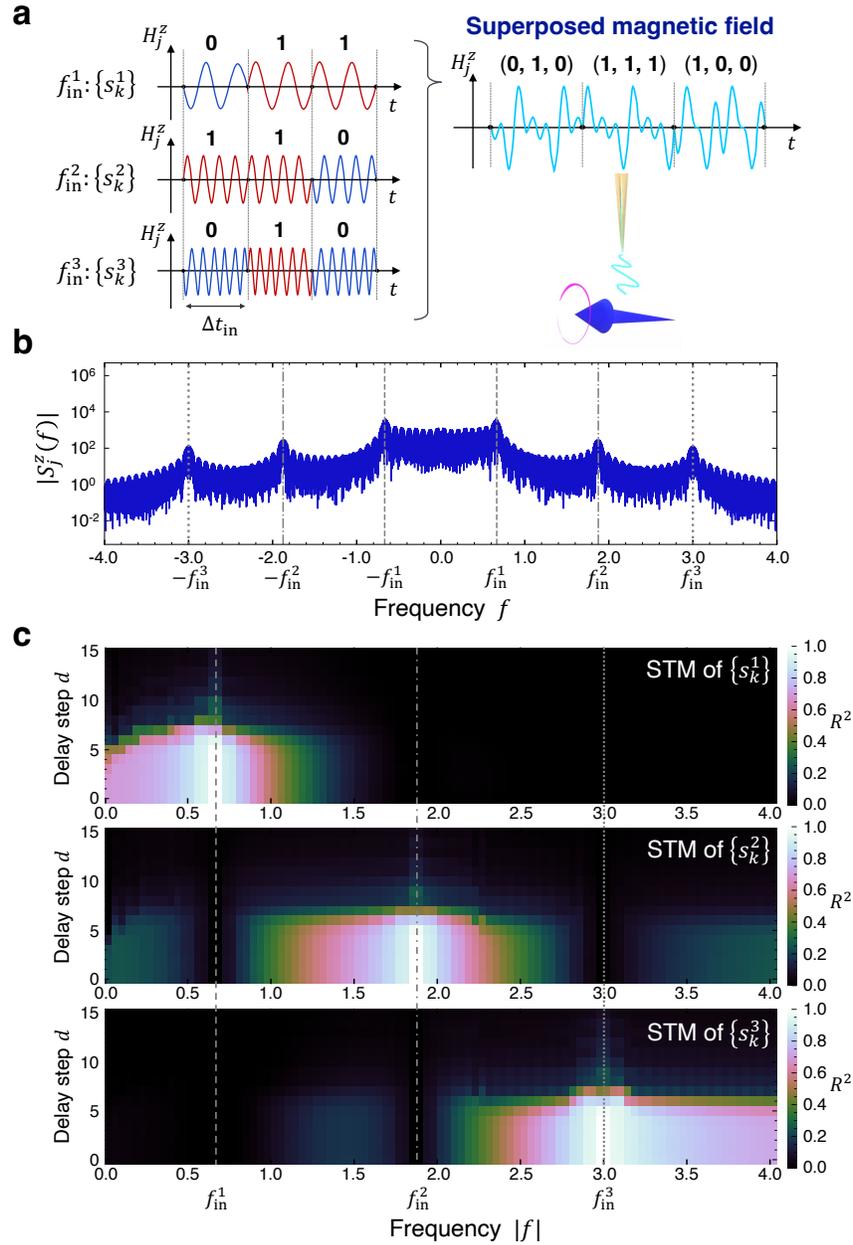

**Fig. 3: Parallel reservoir computing by frequency division multiplexing with the superposed input magnetic fields. a** Schematic of the preparation of input magnetic field. Multiple bit sequences are converted to magnetic fields at different frequencies, and their superposition is given as the input magnetic field. **b** Fourier spectrum of the spin dynamics

$S_j^z(t)$ on the terminal site. The gray lines display the frequencies of the input field: $f_{in}^1 = 8/t_{in}$ for the bit sequence $\{s_k^1\}$ (dashed), $f_{in}^2 = 45/(2t_{in})$ for $\{s_k^2\}$ (dashed-dotted), and $f_{in}^3 = 36/t_{in}$ for $\{s_k^3\}$ (dotted). Strong intensities are observed at around these three input frequencies. **c** Reservoir performance for the STM task in frequency domain. The color represents $R^2$ for the STM task of the input sequence $\{s_k^1\}$ (top), $\{s_k^2\}$ (middle), and $\{s_k^3\}$ (bottom). The gray lines represent $f_{in}^1$, $f_{in}^2$, and $f_{in}^3$ as in **b**. Each STM is retained at around the corresponding input frequency, which allows for multiplexing in frequency domain.

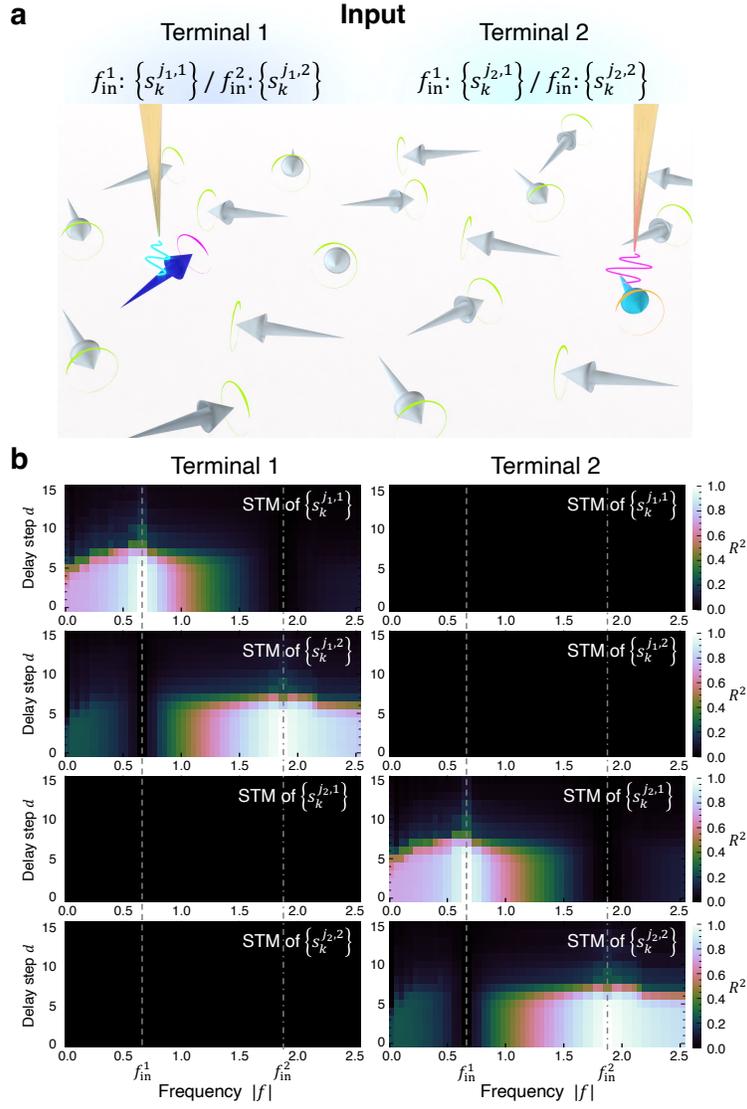

**Fig. 4: Parallel reservoir computing with multiple input/output spin terminals. a** Set up of two input terminal spins. A bit sequence $\{s_k^{j_1,1}\}$ ($\{s_k^{j_1,2}\}$) is given to terminal 1 (the blue arrow) and $\{s_k^{j_2,1}\}$ ($\{s_k^{j_2,2}\}$) is given to terminal 2 (the sky blue arrow) through the magnetic field at the frequency $f_{in}^1 = 8/t_{in}$ ($f_{in}^2 = 45/(2t_{in})$). The yellow needles indicate probes to apply magnetic fields. **b** Reservoir performance for the STM task by independently processing the signals from terminal 1 (left) and terminal 2 (right). The top

two figures represent the STM of $\{s_k^{j_1,1}\}$ and $\{s_k^{j_1,2}\}$, and the bottom two figures show the STM of $\{s_k^{j_2,1}\}$ and $\{s_k^{j_2,2}\}$. The gray lines correspond to $f_{\text{in}}^1$ and $f_{\text{in}}^2$. The STM of $\{s_k^{j_1,1}\}$ and $\{s_k^{j_1,2}\}$ can be recovered only by the signals from the terminal 1, and the STM of $\{s_k^{j_2,1}\}$ and $\{s_k^{j_2,2}\}$ requires the signals from the terminal 2 for reproduction. Each memory is retained at around the corresponding input frequency. A straightforward extension of this scheme would allow for high-density integration of computational units in both spatial and frequency domains.

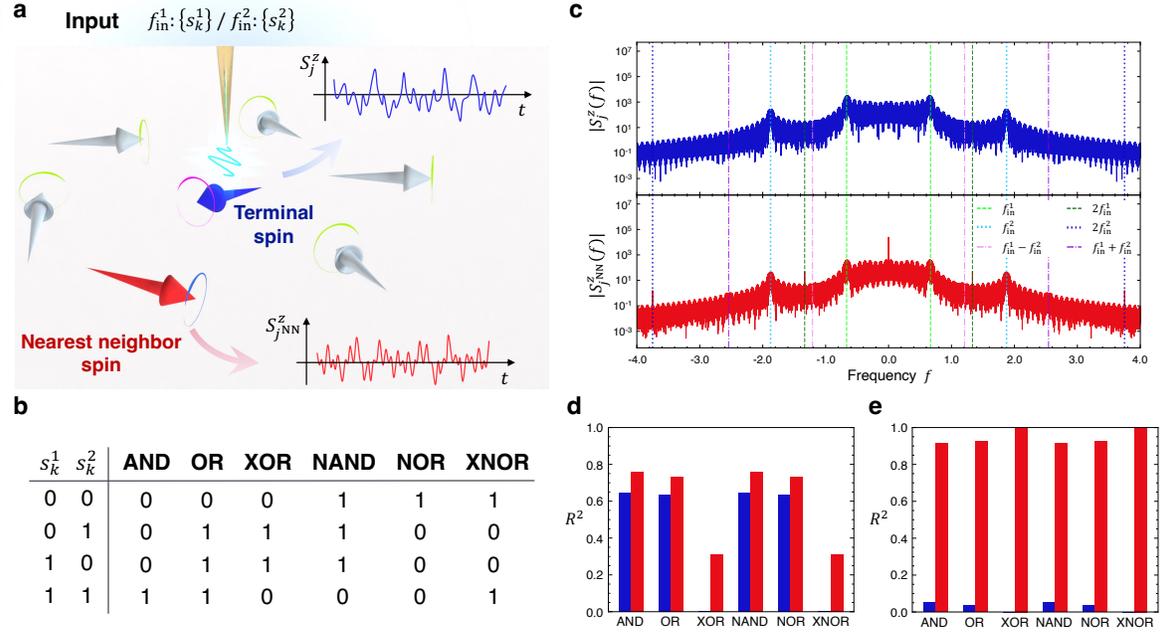

**Fig. 5: Logic gate operations of two input bits stored in different frequencies. a** Schematic of the input terminal spin (the blue arrow) and its nearest neighbor spin (the red arrow). The former is directly affected by the input magnetic field through a probe represented by the yellow needle, while the latter is affected via the exchange interaction between the spins. **b** Truth table for logic gates with input bits $s_k^1$ and $s_k^2$. **c** Fourier spectra of the dynamics of the terminal spin $S_j^z(t)$ (top) and the nearest neighbor spin $S_{j^{NN}}^z(t)$ (bottom). The dashed lines display $f_{in}^1 = 8/t_{in}$ (light green) for the bit sequence $\{s_k^1\}$ and $2f_{in}^1$ (dark green). The dotted lines show $f_{in}^2 = 45/(2t_{in})$ (light blue) for the input sequence $\{s_k^2\}$ and $2f_{in}^2$ (dark blue). The dashed-dotted lines represent the second harmonics frequencies $f_{in}^2 - f_{in}^1$ (pink) and $f_{in}^1 + f_{in}^2$ (purple). Strong intensities appear at around $\pm f_{in}^1$ and $\pm f_{in}^2$ in both spectra, while only that of the nearest neighbor spin shows peaks at $\pm 2f_{in}^1$, $\pm 2f_{in}^2$, $\pm(f_{in}^2 - f_{in}^1)$ and $\pm(f_{in}^1 + f_{in}^2)$. **d** Reservoir performance for the logic gate tasks with frequency filters centered on $\pm f_{in}^1$ and $\pm f_{in}^2$. The blue and red bars represent $R^2$ utilizing the dynamics of the terminal spin and the nearest neighbor spin, respectively. The linearly

inseparable gates, XOR and XNOR, are nearly impossible with these frequency filters. **e** The same plot as **d** with frequency filters centered on $\pm(f_{in}^2 - f_{in}^1)$ and $\pm(f_{in}^1 + f_{in}^2)$. The performance of the nearest neighbor spin becomes much better for all gating operations including the inseparable gates, owing to the nonlinear processes via the exchange interaction.

Supplementary Information for

# Thermally-robust spatiotemporal parallel reservoir computing by frequency filtering in frustrated magnets

Kaito Kobayashi and Yukitoshi Motome

*Corresponding author. E-mail: kaito-kobayashi92@g.ecc.u-tokyo.ac.jp

**Supplementary Note 1: Thermal robustness for larger input magnetic fields**

In Supplementary Fig. 1, we show the Fourier spectra of the spin dynamics and performance for the STM task when the amplitude of the input magnetic field $H_{\text{in}}$ is increased from 0.1, used in the main text, to 0.15 and 0.2. The peak intensity near the frequency $f_{\text{in}}$ increases as $H_{\text{in}}$ is amplified, greatly exceeding the noise level of thermal fluctuations. Consequently, $R^2$ around $f_{\text{in}}$ is improved, especially for $T = 0.1$, indicating an extension of the operational temperature range.

**Supplementary Note 2: Optimal window of frequency filters**

To investigate the information distributions in frequency domain, we simultaneously utilize two frequency windows of $(m_1 - 1)/(2at_{\text{in}}) \leq |f| < m_1/(2at_{\text{in}})$ and $(m_2 - 1)/(2at_{\text{in}}) \leq |f| < m_2/(2at_{\text{in}})$ with positive integers $m_1, m_2$ and $a = 4, 8$. Supplementary Figure 2 represents the performance for the STM task of delay $d = 0$ while varying the combination of two frequency windows. The $R^2$ on the diagonal line $m_1 = m_2$, which corresponds to the performance with a single frequency window, is relatively low compared to $R^2$ with two windows, $m_1 \neq m_2$, demonstrating the performance improvement by utilizing the spin dynamics in a wider frequency range. However, if two frequency windows are mutually linked by the specific unit of $1/t_{in}$, the performance remains comparable to that obtained with a single window because almost the same information is retained in these frequencies. This is evidenced by the relatively darker lines in Fig. S2 on $m_2/(2at_{in}) = m_1/(2at_{in}) \pm n/t_{in}$ and $m_2/(2at_{in}) = n/t_{in} - (m_1 - 1)/(2at_{in})$ with a positive integer $n$, which signifies the equivalence of information in $f$ and $f \pm n/t_{in}$, and $f$ and $n/t_{in} - f$, respectively. Therefore, the minimum bandwidth required to access all the information is $1/(2t_{in})$, and the frequency window of $(m-1)/(2t_{in}) \leq |f| < m/(2t_{in})$ adopted in the main text is optimal as it avoids the inclusion of overlapped information.

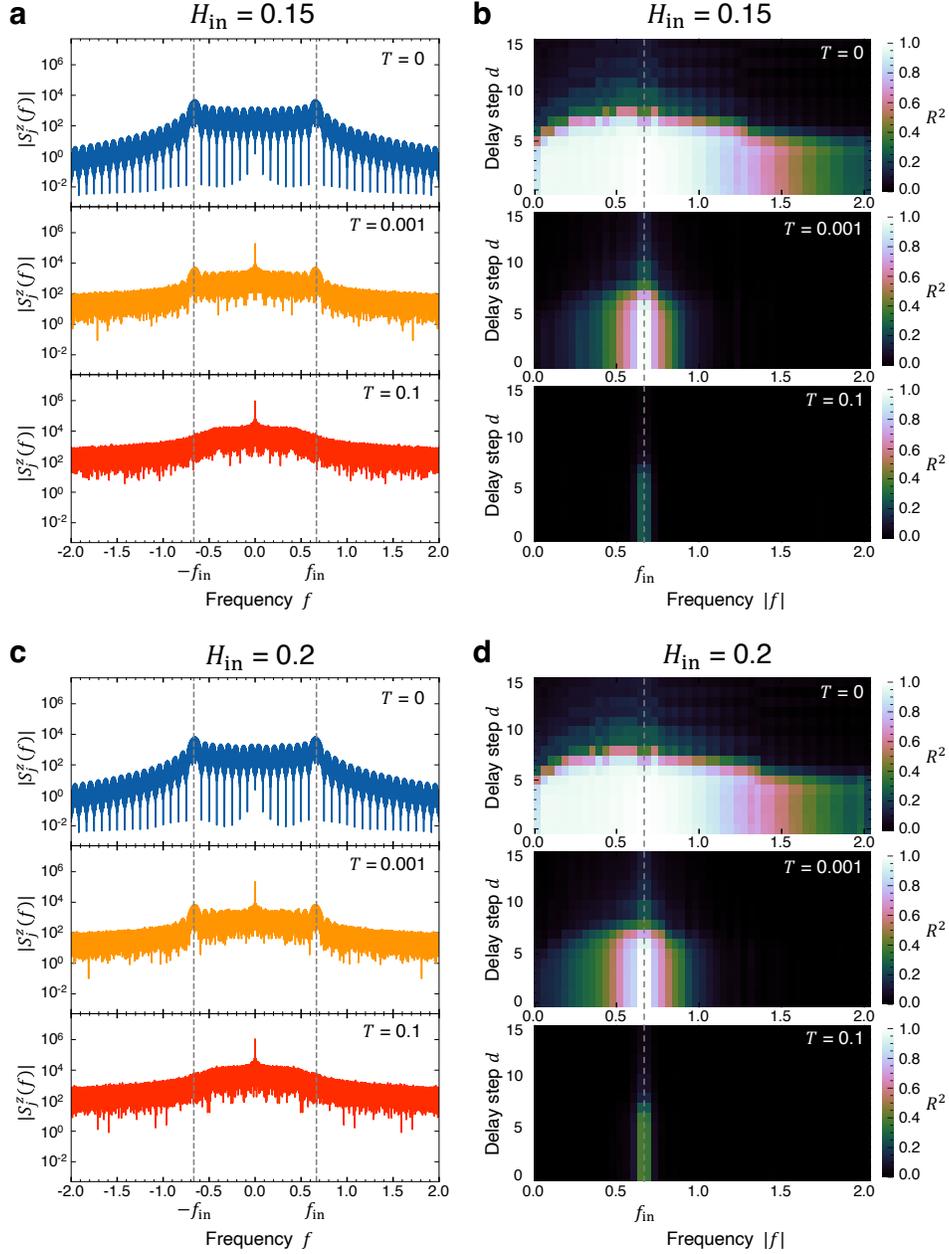

**Supplementary Fig.1: Reservoir computing at finite temperature with stronger input magnetic fields. a** and **c** Fourier spectra of the spin dynamics on the terminal site at three different temperatures. The dashed lines display the input frequency $f_{in} = 8/t_{in}$. **b** and **d** Distributions of the STM in frequency domain. The color represents the value of $R^2$ and the dashed line shows $f_{in}$. The amplitude of input magnetic field is **a** and **b** $H_{in} = 0.15$ and **c** and **d** $H_{in} = 0.2$.

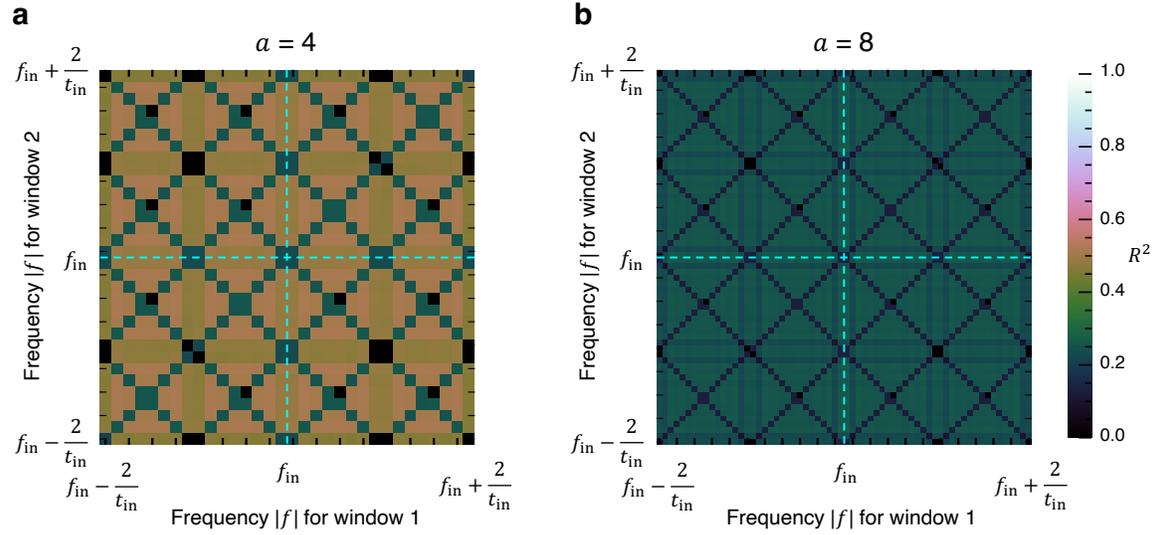

**Supplementary Fig. 2: Performance of the STM task with two frequency windows. a** and **b** The color plot of $R^2$ for the STM task of delay step $d = 0$ at zero temperature. The horizontal and vertical axes represent the frequency for the window 1 and 2, respectively. The frequency window 1 passes signals within **a** $(m_1 - 1)/(8t_{in}) \leq |f| < m_1/(8t_{in})$, **b** $(m_1 - 1)/(16t_{in}) \leq |f| < m_1/(16t_{in})$ with a positive integer $m_1$. The frequency window 2 similarly transmits signals within the same frequency range with a positive integer $m_2$ instead.